\documentclass[11pt]{article}
\usepackage[utf8]{inputenc}
\usepackage{amsmath}
\usepackage{amsfonts}
\usepackage{amssymb}
\usepackage{graphicx}
\usepackage[margin=1in]{geometry}
\usepackage{natbib}
\usepackage[affil-it]{authblk}
\usepackage{color}
\usepackage{hyperref}
\usepackage[cal=cm]{mathalfa}
\usepackage{setspace}
\newcommand{\bfalpha} {\boldsymbol{\alpha}}
\newcommand{\bfbeta} {\boldsymbol{\beta}}
\newcommand{\bfgamma} {\boldsymbol{\gamma}}

\newcommand{\bfxi} {\boldsymbol{\xi}}
\newcommand{\bfXi} {\boldsymbol{\Xi}}
\newcommand{\bfmu} {\boldsymbol{\mu}}

\newcommand{\bftheta} {\boldsymbol{\theta}}

\newcommand{\bfnu}{\boldsymbol{\nu}}

\newcommand{\bfY} {\mathbf{Y}}
\newcommand{\bfX} {\mathbf{X}}

\newcommand{\bfN} {\mathbf{N}}

\newcommand{\bfy} {\mathbf{y}}

\newcommand{\bfSigma} {\boldsymbol{\Sigma}}
\newcommand{\E}{\mathsf{E}}

\newcommand{\vecc}{\mathsf{vec}}
\newcommand{\trace}{\mathsf{tr}}
\newcommand{\normal}{\mathsf{N}}

\newcommand{\MN}{\mathsf{MN}}
\newcommand{\Poi}{\mathsf{Poi}}

\providecommand{\keywords}[1]{\textbf{\textbf{Keywords:}} #1}

\graphicspath{{./Figures/}}

\title{A stratified age-period-cohort model for spatial heterogeneity in all-cause mortality}

\author{Theresa Smith\thanks{Department of Mathematical Sciences, University of Bath, \texttt{T.R.Smith@bath.ac.uk}}}
\date{}

\begin{document}

\begin{singlespace}
   
\maketitle

\begin{abstract}
A common goal in modeling demographic rates is to compare two or 
more groups. For example comparing mortality rates between men and women or between geographic regions may reveal health inequalities. 
A popular class of models for all-cause mortality as well as incidence of specific diseases like cancer is the age-period-cohort (APC)
model. Extending this model to the multivariate setting is 
not straightforward because the univariate APC model suffers from well-known identifiability problems. Often APC models are fit separately for each strata, and then comparisons are made post hoc. 
This paper introduces a stratified APC model to directly assess 
the sources of heterogeneity in mortality 
rates using a Bayesian hierarchical model with matrix-normal priors that share information on linear and nonlinear aspects of the APC effects across strata.  Computing, model selection, and prior specification are addressed and the model is then applied to all-cause mortality data from the European Union.\\

\noindent \keywords{Bayesian age-period-cohort analysis, spatio-temporal models, human mortality database, demographic statistics}
\end{abstract}
\end{singlespace}
%
%

\section{Introduction and Running Example} \label{S1}

Age-period-cohort (APC) models are used to model demographic rates 
including
all-cause mortality as well as incidence of specific diseases like 
cancer. 
APC analysis is conceptually attractive because the models capture the 
major time scales along which disease rates evolve where
direct measurements of the disease generating processes at the 
population scale are difficult or impossible to gather.
Age effects are a surrogate for cumulative wear and tear on the 
body, period (time of event) effects for short acting exposures or new treatments, 
and cohort (time of birth) effects for longer acting exposures such as smoking.
APC models are used to smooth out time trends in disease, forecast future numbers of cases, and give
clues for the most important etiological factors.
However they suffer from a well known identifiability issue
wherein the linear trends along these time scales cannot be disentangled 
without adding additional constraints.

A common goal in applications of APC models is to compare disease rates
across groups such as sexes, disease subtypes, or regions.  
Often this is done by post-hoc comparisons of separate APC
models. 
In a recent example, \cite{wang:etal:2018} fit separate models to drowning 
rates for males and females in China using the APC web-tool 
from the US National Institutes of Health 
\citep{rosenberg:etal:2014} and then compared graphically. The dominant approach when multivariate APC analyses are done is based on multivariate Gaussian Markov random fields (GMRFs)
\citep{riebler:held:10,riebler:etal:12, papoila:etal:2014}. 
Typical implementations of univariate GMRFs include linear constraints, which then
mask identifiability issues in multivariate extensions. 
 
We resolve these issues with a novel stratified APC model based on non-linear aspects of the age, period, and cohort effects that are identifiable. We previously considered this parameterization in a Bayesian context for univariate APC models in \citep{smith:wakefield:16}.  A Bayesian hierarchical model with matrix-normal priors allows for pooling of information on these identifiable terms across strata, and we carefully consider the specification of cross-correlation structures and prior distributions. Practical issues such as computing with \texttt{r-INLA} and model selection are also discussed. 

\begin{figure}
\includegraphics[width = \textwidth, trim={0 1mm 0 1mm},clip]{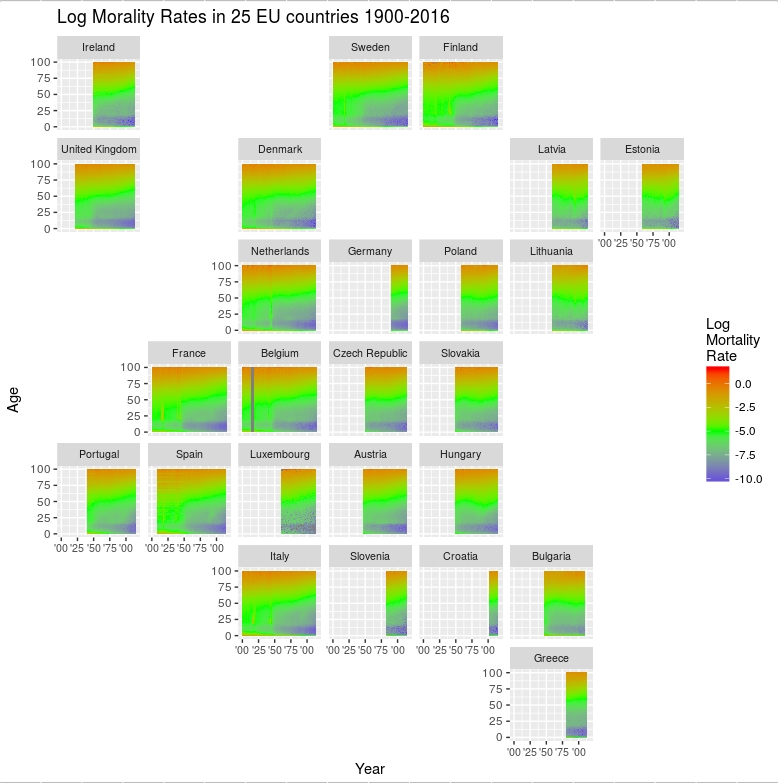}
\caption{Single age-year log all cause mortality for 1900--2016, ages 0--100
in 25 EU member countries.} \label{logEUmort}
\end{figure}

We apply this model to mortality rates in 25 European Union (EU) countries 
from the Human Mortality Database (HMD). The HMD contains 
raw data on births and deaths reported by the national statistics 
offices of 39 countries as well as computed mortality rates and life 
tables.  While not directly shareable, the data are free to download 
after registration (\url{http://www.mortality.org}).
Figure \ref{logEUmort} shows the single age-year death rates on the log 
scale for 25 EU countries. Cyprus, Malta and Romania are not covered by 
the HMD. Several interesting features of this data are immediately 
apparent.  First, the mortality series are of very different lengths.  
While some countries in northern and western Europe (e.g., Denmark or 
France) have very long series dating back even before 1900, a common start 
year for data from countries in eastern Europe is 1959, the year of the 
first Soviet Census after World War II \citep{schwartz:86}. Some of the shorter series (e.g., 
Croatia or Ireland) are related to countries becoming independent or 
in the case of Germany, reunified. 

There are clear similarities in the mortality trends. Early childhood 
mortality rates are decreasing, and most countries show improvement in 
mortality rates for the middle aged. The toll of the two world wars is pronounced, with bright bands of higher mortality 
across all age groups in several countries. There is a general 
shift towards lower mortality in working years after WWII with wider 
understanding and usage of vaccines and antibiotics as well as further 
development of state-based public health initiatives. The changes in mortality rates also show some  geographic patterns.  
 For example, the bright green frontier of higher  mortality shows a 
 consistent change from ages 50 to 75 in northern, southern, 
and western Europe but is relatively flat in eastern Europe. 

The remainder of this paper is as follows. In Section \ref{S2} we review the identification problem in univariate APC models and 
discuss stratified APC models
based on multivariate GMRFs. 
The new  stratified APC model is introduced in Section \ref{S3}, with implementation details in Section \ref{S4}. We conclude with the application to all-cause mortality in the EU in Section \ref{secHMD} and a discussion
of other research directions.

\section{Background} \label{S2} 
\subsection{Univariate APC models}\label{s22}
Let $\bfY = \{y_{ij}:  i = 1, \dots, A;  j= 1, \dots , T\}$ be a matrix of the
 number of cases in each of $A$ age groups and $T$ periods, and let $\mathbf{N} = \{N_{ij}, i = 1, \dots, A; j = 1, \dots, T\}$ be the number of person-years at risk in each age group and time period.  The cohort index
  $k$ is a function of age and period. When the age scale and time scale are the 
  same (e.g., 5 year age groups and 5 year time intervals), then $k = A-i +j$ 
 so that $k \in \{1, \dots, A+T - 1\}$.  
 In this paper we assume $y_{ij} \sim \Poi(N_{ij} \exp \mu_{ij})$, but 
the methods described can easily be used for other likelihoods. For example, for country-level all-cause mortality it may be reasonable to assume a Gaussian likelihood directly on the log mortality rates. 
The basic  APC model for the log rate is \citep{clayton:schifflers:87a}
\begin{align} \label{basicapc}
\log\E \left[ \frac{y_{ij}}  {N_{ij}} \right] & = \mu_{ij} = \delta + \alpha_i + \beta_j + \gamma_{k}.
\end{align}
It is well known that the parameters in (\ref{basicapc}) are not directly interpretable
as log relative risks because the model is over parametrized. 
\citet{kuang:etal:08identification} and \citet{nielsen:etal:14}, following \cite{carstensen:07},  define this identifiability issue
from a group theoretic perspective.   The overall mean, as given in
(\ref{basicapc}), is invariant to a translation on each set of effects
and the addition of a linear trend in the age, period, and cohort
parameters. 
Let $\mathcal{G} = \{g: g\bftheta = (g\delta, g\bfalpha, g\bfbeta, g\bfgamma) \}$ be a group of transformations where 
\begin{eqnarray}
g \delta &= & \delta -a -b -c + (A-1)d\label{eq:g1}\\
 g \bfalpha& =& \left\{\alpha_i  + a + (i-1)d \right\}_{i = 1}^{A} \label{eq:g2}\\
g\bfbeta &=  & \left\{\beta_j  +b - (j-1)d \right\}_{j = 1}^T \label{eq:g3}\\
g\bfgamma& = &\left\{\gamma_{k = A-i +j } + c + (A- i+j -1)d \right\}_{i=1,j=1}^{i = A, j = T}\label{eq:g4}
\end{eqnarray}
for any real numbers $a,b,c,$ and $d$.  An interpretation of these numbers is that $a$, $b$, and $c$ are the overall levels of the age, period, cohort effects, respectively, and $d$ is the linear trend.
The log rates are invariant with respect to these transformations.  That is, for any $g \in \mathcal{G}$, 
$\mu_{ij}(g\bftheta) = \mu_{ij} (g\delta, g\alpha_i, g\beta_j, g\gamma_k) = \mu_{ij}(\delta, \alpha_i, \beta_j, \gamma_k) = \mu_{ij}(\bftheta)$:
\begin{align*}
 \mu_{ij} (g \delta, g\alpha_i, g\beta_j, g\gamma_k)
& = \left(\delta -a -b -c - (A-1)d\right) + \left(\alpha_i  + a + (i-1)d\right)  \\
& \, +\left ( \beta_j  +b - (j-1)d \right) + \left(\gamma_{k} + c+ (A - i +j -1)d \right) \\
& = \delta + \alpha_i + \beta_j + \gamma_{k}\\
& = \mu_{ij}(\delta, \alpha_i, \beta_j, \gamma_k).
\end{align*}
Typically the distribution of the observed data only depends on the age, period, and cohort
parameters through the log rates.
Thus the likelihood itself is invariant with respect to $\mathcal{G}$, and the full set of age, period, and cohort effects is not identifiable.  More specifically, the overall 
level and linear trend in each set of effects are unidentifiable.

The issue of overall levels can be resolved uncontroversially in the univariate case 
with sum-to-zero constraints on the $\bfalpha$, $\bfbeta$ and $\bfgamma$ terms or by 
setting one category from each of these terms as the reference category. 
The second form  of non identifiability arises because of the linear 
dependence between the age, period, and cohort indices, and there is no genuine 
solution to this problem though many in the literature have proposed adding constraints based on expert knowledge or convenient mathematical criteria \citep{fienberg:13}. 
However, seemingly innocuous 
differences in the choice of such constraints can yield contradictory estimates of $
\bfalpha$, $\bfbeta$, and $\bfgamma$. 

\subsection{Existing stratified APC models} \label{sec22}

An important extension of the basic APC model is to incorporate time-independent factors (strata) such as gender, region, or disease
subtype.   Let $r, r = 1,\dots, R$ represent  the index for strata,  then
  the most general stratified APC model is
  $$ \mu_{ijr} = \delta_r + \alpha_{ir} + \beta_{jr} + \gamma_{kr},$$
so that the full parameter space $\boldsymbol{\Theta}$ is an $R \times 2(A + T)$ matrix with rows
$(\delta_r, \bfalpha_r , \bfbeta_r, \bfgamma_r)$, $r = 1,\dots, R$.
One can separately apply a stratum-specific $g \in \mathcal{G}$ with $(a_r, b_r, c_r,d_r )$ to 
each row of $\boldsymbol{\Theta}$ without changing the log rates.  Hence, $4R$ constraints 
are required to produce identifiable age, period, and cohort effects. 

In a stratified APC model, certain cross-strata 
   relative risks are identifiable up to a multiplicative constant if one set of effects is shared across strata.
 Specifically, for the other two sets of non-shared effects, trends in relative risks within the same time index but between strata are identifiable.
 For example, suppose the age effects are shared across strata (that is $\alpha_{ir}=\alpha_i$ for all $i$), then the unidentifiable
 level of the age effects and the unidentifiable linear trend are common to all strata: $a_r =a$ and $d_r = d$. The relative risk between strata $r_1$ and $r_2$ in time period $1$ is  
$\exp(\beta_{1r_1} - \beta_{1r_2})$. 
For two transformations defined by $(a, b_{r_1}, c_{r_1},d)$  and $(a, b_{r_2}, c_{r_2},d)$ this becomes 
\begin{align*}
g_1 \beta_{jr_1} - g_2 \beta_{jr_2} &= \beta_{jr_1} - \beta_{jr_2} + (b_{r_1} - b_{r_2})\nonumber\\
\left[\exp(g_1 \beta_{jr_1} - g_2 \beta_{jr_2})\right] &= \exp(\beta_{jr_1} - \beta_{jr_2})\exp(b_{r_1} -b_{r_2}).
\end{align*} 
Thus, in a plot of $\exp(\beta_{jr_1} - \beta_{jr_2})$ against $j$, the y-axis scaling is ambiguous (i.e.,~can be multiplied by any positive number), but the overall shape--including the existence and direction of linear trends--can
be ascertained.   Note that all of the identifiability problems discussed in Section \ref{s22} still persist within each stratum.

In \cite{riebler:held:10} and \cite{riebler:etal:12}, 
the age effects are assumed to be the same across regions; in \cite{riebler:etal:12suicide}, 
period effects were shared, again across regions; and in \cite{papoila:etal:2014}, cohort effects are share by gender. For example, the model in \cite{riebler:etal:12} is
$$ \mu_{ijr} = \delta_r + \alpha_{i} + \beta_{jr} + \gamma_{kr}.$$
The authors fit these models in a Bayesian framework using correlated intrinsic 
Gaussian Markov random fields priors on each set of time effects. Specifically the prior on the age effects and the marginal prior on the period and cohort effects for a given region is a second order random walk (RW-2) prior, and the cross correlation between the effects of 
different regions is incorporated using a Kronecker-structured covariance. 

With a single, univariate intrinsic GMRF prior, sum-to-zero constraints are typical because the precision matrix is not full rank \citep{rue:held:2005}, and in the GMRF-based multivariate APC models, there are many sum-to-zero constraints.
Thus these multivariate RW-2 models seem avoid the issue of multiplicative ambiguity in the cross-strata relative risks, when in fact they are only identifiable conditional on 
an implicit choice of $b_r$'s and $c_r$'s made by these constraints. 
While the sum-to-zero constraints were uncontroversial in the univariate APC 
model, they can give the misleading impression that these cross-strata relative 
risks are fully identifiable.



\section{Methodology} \label{S3}

\cite{smith:wakefield:16} explored Bayesian inference for an APC parameterization based on non-linear aspects of the age, period, and cohort effects, and in this section we review this model and extend to the hierarchical setting using matrix-normal priors. 
\subsection{Canonical parameterisation} \label{ss:mnn}
\citet{kuang:etal:08identification},
 \cite{nielsen:etal:14}, and \cite{miranda2014inference} put forward a 
 univariate APC model for data with equal-width age and time intervals in terms of three initial log rates and a full set of second differences in the age, period, and cohort effects. There exists a unique, full rank design matrix, $M$, such that the vector of log rates $\bfmu$ can be expressed as
\begin{align} \label{mnn1}
\bfmu = M\bfxi, \, \xi = \{\mu_{{ijk}_1}, \mu_{{ijk}_2}, \mu_{{ijk}_3}, \Delta^2 \bfalpha, \Delta^2 \bfbeta, \Delta^2 \bfgamma\}.
\end{align}
Here $\bfxi$ is a vector of length $2(A+T) - 4$, which is exactly the number of parameters in (\ref{basicapc}) less the number of free constants defining the group $\mathcal{G}$. 

The entries in $M$ are determined by the choice of coordinates for the base rates $\mu_{{ijk}_1}, 
\mu_{{ijk}_2}, \mu_{{ijk}_3}$. \citet{kuang:etal:08identification} and  
\cite{nielsen:etal:14} chose three initial time points using age-cohort indices (i.e., $\mu_{ijk} 
= \mu_{ik}$), and \cite{miranda2014inference} use age-period indexing (i.e., $\mu_{ijk} = \mu_{ij}
$). The three pair of indices must be chosen so that they define a triangle rather than a line. So 
$\{\mu_{11}, \mu_{12}, \mu_{13}\}$ is not an acceptable choice, but $\{\mu_{11}, \mu_{12}, \mu_{21}\}$ 
is in either coordinate system.  For simplicity, we assume this requirement is satisfied and use the more general notation $\{\mu_{{ijk}_1}, \mu_{{ijk}_2}, \mu_{{ijk}_3}\} =\{\mu_0^1, \mu_0^2, \mu_0^3\}$.

If we choose the largest or smallest values of $i$ and $j$ or $k$, the baseline rates are in the corners of the data array. 
For example, \cite{miranda2014inference} use the earliest periods and the oldest age groups:
$$\{\mu_0^1, \mu_0^2, \mu_0^3\} = \{\mu_{A1}, \mu_{A-1,1}, \mu_{A2}\}.$$
Alternatively, we can choose values in the middle of the array.  Suppose $A$ is odd and let $U = (A+1)/2$ be the middle age index.  \cite{nielsen:2015} suggests using 
$$\{\mu_0^1, \mu_0^2, \mu_0^3\} = \{\mu_{UU}, \mu_{U+1,U}, \mu_{U,U+1}\}$$
in age-cohort coordinates.  
A further option is to parametrize in terms of one initial point and two differences:
\begin{align}\label{mnn2}
\bfmu = M^\star \bfxi^\star, \, \bfxi^\star = \{\mu_0^1, \mu_0^2-\mu_0^1, \mu_0^3-\mu_0^1, \Delta^2 \bfalpha, \Delta^2 \bfbeta, \Delta^2 \bfgamma\},
\end{align}
where `$\star$' emphasizes that the design matrix in \ref{mnn1} and \ref{mnn2} are different in the first three columns, but the estimates of the second differences are identical: $\bfxi[4:(2(A+T) - 4)] = \bfxi^\star[4:(2(A+T) - 4)]$.  An example design matrix is in the \hyperref[supps]{Supplementary Materials}.

The first three terms in this parametrization define a simple 
linear model for changes in disease (log) rates over time with an intercept ($
\mu_0^1$) and two slopes ($\mu_0^2-\mu_0^1$ and $\mu_0^3-
\mu_0^1$). The remaining terms then describe the non-linear 
deviations from this simple specification. For example, assuming $\{\mu_0^1, \mu_0^2, \mu_0^3\} = \{\mu_{UU}, \mu_{U,U+1}, \mu_{U+1,U}\}$, the log rates in age-cohort indices are
\begin{align} \label{ACmid}
\mu_{ik} =& \mu_{UU} + (i-U)(\mu_{U+1,U} - \mu_{UU}) + (k-U)(\mu_{U,U+1} - \mu_{UU}) \\
& \qquad \qquad + M_{[,-(1:3)]}\bfxi_{[-(1:3)]}. \nonumber
\end{align}
This is similar to the 
models discussed in \citep{rosenberg2011age}, which include an intercept,
two slopes (net drift and either the longitudinal or cross-sectional age trend), and non linear terms.
The curvature terms (second differences) can also be interpreted as accelerations in the trends along the 
three time scales \citep{clayton:schifflers:87a}.

\subsection{Matrix variate normal priors for multi-way data} \label{sec:mvn}

\cite{smith:wakefield:16} introduced a Bayesian hierarchical model 
based on the parameterization in Section \ref{ss:mnn} by incorporating 
uniform priors on the three initial points and Gaussian priors on the 
vectors of second differences.  The main benefit of Bayesian inference 
in that paper was to shrink the curvature terms (in particular the 
cohort terms) that are not be well estimated by maximum likelihood. 
Here we extend this model by using matrix-normal priors, which will 
again provide some penalization but also allow for pooling of 
information across strata.

The matrix normal prior is a common tool for analyzing multi-way data  including multivariate spatial data \citep{gelfand:vounatsou:03, knorrheld:00, smith:etal:15} as well as mortality data \citep{fosdick:hoff:14}. 
 Let $\bfX\sim\MN(\bfnu, \bfSigma_C, \bfSigma_R)$ denote the matrix normal distribution with separable covariance structure and mean $\bfnu$ \citep{dawid:81}. This means $\vecc\left( {\bfX}\right) \sim \normal\left(\vecc(\bfnu),\bfSigma_{C} \otimes \bfSigma_{R}\right)$, where `$\vecc(\cdot)$' stacks the columns of a matrix into a single vector and  `$\otimes$' is the Kronecker product. Here $\bfSigma_C$ is the column covariance matrix describing correlation arising from the feature indexing the columns (e.g., the strata), and  similarly the row covariance matrix $\bfSigma_R$ describes the  correlation arising the feature indexing the rows. A mean of $\boldsymbol{\nu}=0$ is typical when using the matrix normal distribution as a prior for latent random effects.

The separable covariance structure in the matrix normal is computationally convenient because 
only the individual row and column covariance matrices and never the full covariance matrix appear in calculations.  Suppose $n_r$ and $n_c$ are the row and column dimensions, then the 
probability density function for the matrix normal can be written as
\begin{align*}
&\left(2\pi|\bfSigma_{C} \otimes \bfSigma_{R}|\right)^{-1/2}\exp\left\{-\vecc(\bfX-\bfnu)' (\bfSigma_{C} \otimes \bfSigma_{R})^{-1}\vecc(\bfX-\bfnu)/2 \right\}\\
& = \left(2\pi|\bfSigma_{C}|^{n_r} |\bfSigma_{R}|^{n_c}\right)^{-1/2}
\exp\left\{-\trace\left[\Sigma_C^{-1}(\bfX-\bfnu)'\Sigma_R^{-1}(\bfX-\bfnu) \right]  /2\right\}.
\end{align*}

\subsection{An identifiable stratified APC model} \label{sapc}
Let the vectors $\bfy_r$ and $\bfN_r$, $r =1, \dots R$ represent the numbers of deaths and population sizes for the $r^{\text{th}}$ stratum. We assume the counts follow a Poisson distribution with rates broken into the age, period, and cohort effects according to the
identifiable parameterization in \ref{ss:mnn}: 
\begin{align*}
\bfy_{r} & \sim \mathsf{Poi}(\bfN_{ijr} \exp(M \bfxi_r))\\
  \bfxi_r & =\{\mu_{0r}^1, \mu_{0r}^2, \mu_{0r}^3,\Delta^2 \bfalpha_r,\Delta^2 \bfbeta_r,\Delta^2 \bfgamma_r \}\\
& \equiv \{  \bfxi_{0r} , \bfxi_{\alpha r},\bfxi_{\beta r},\bfxi_{\gamma r}  \}
\end{align*}
We organize the second differences and baseline rates into matrices.  For example, we combine the $\bfxi_{\alpha r}$'s to get an $(A-2)  \times R$ matrix:  $\bfXi_\alpha$ where
$$\bfXi_{\alpha [ir]} = \Delta^2 \alpha_{i+2,r}$$
and similarly define $\bfXi_{0}, \bfXi_{\beta}, \bfXi_{\gamma}$.
We use the matrix normal prior to allow for correlation between the accelerations across strata:
\begin{align*}
\bfXi_\phi &\sim \mathsf{MN}\left(0; \tau^{-1}_\phi \mathcal{I}, \Sigma_s(\rho_\phi)\right), \quad \phi = \alpha, \beta, \gamma.
\end{align*}
The first covariance matrix is the scaled  identity matrix, 
meaning that, within a  stratum, the second differences or
log baseline rates are independent. The second 
variance matrix encodes the cross-strata correlation, the 
strength of which may be governed by a parameter $\rho$.  Note 
that we cannot have a scale parameter for both covariance 
matrices because only the product of these two is 
identifiable.   In this model, the separable covariance structure means assuming that the 
covariance in the curvatures factors into a purely between-strata portion and a purely 
between-time portion.
For the baseline points, we still use a matrix normal prior but include the mean as a parameter instead of assuming a zero mean:
\begin{align*}
\bfXi_0 &\sim \mathsf{MN}\left(\bfnu_0; \tau^{-1}_0 \mathcal{I}, \Sigma_s(\rho_0)\right).
\end{align*}

Three possible choices for $\Sigma_s$ are considered here: independent, exchangeable, and spatial. 
The first two are straightforward.  For independence we have $\Sigma_s = \mathcal{I}$ and for the exchangeable model 
$$ \Sigma_s[ij] = \begin{cases}
1& i = j,\\
\rho & i \neq j
\end{cases}\quad  \rho \in \left(-\frac{1}{R-1}, 1\right).
$$
Where the strata represent geographic regions, a covariance 
matrix encoding larger correlations for near-by areas and 
smaller correlations for areas far away from each other is of 
interest. A classical choice is to base this around the neighborhood structure of the regions via an adjacency matrix $W$: 
$$ W_{ij} = \begin{cases}
1, & j \in \text{nbhd}(i)\\
0, &j \not\in \text{nbhd}(i).
\end{cases}
$$
where  $j \in \text{nbhd}(i)$  means areas $i$ and $j$ share a border. This forms the basis of the intrinsic conditional-autoregressive (ICAR) model with precision matrix
$Q_s= D -W,$
where D is a diagonal matrix with entries equal to 
the row sums of W. Often independent random effects 
are used alongside spatial random effects 
to account for both structured and unstructured 
correlation. 

Here we follow the recommendations of 
\cite{riebler:etal:16} and use the so-called BYM2 
model where the random effect is the weighted sum of 
an independent and a spatial component, with the 
spatial component scaled by $\sigma^2_\text{ref}$, the geometric mean of the 
marginal variances under $Q_s$ as defined above.  This 
scaling is important because the prior variance of the
random effects depends on the neighborhood structure in the ICAR prior \citep{sorbye:rue:14}.
Under the BYM2 model, the cross-stratum correlation matrix is
$$\Sigma_s(\rho) = (1-\rho) \mathcal{I} - \rho Q_{\star}^{-}, \quad \rho \in (0,1)$$
where  $Q_{\star}^{-}$ is the generalized inverse of $Q_s$, scaled by 
$\sigma^2_\text{ref}$.
%
An alternative to this areal approach is to construct a spatial covariance matrix based on the distance between some measure of the center of each region, but we do not pursue this option here. 
%
%
%


The stratified APC model proposed here has some similarities with the multivariate RW-2 models discussed
in Section \ref{sec22}.  We can re-write a second-order random walk for regular time intervals as
\begin{align*}
\bfalpha &\sim \text{RW-2}(\tau_\alpha = 1) \\
&\implies
\alpha_i | \bfalpha_{-i} \sim N \left( \frac{2}{3}(\alpha_{i-1}  + \alpha_{i+1})  -\frac{1}{6} (\alpha_{i-2}  + \alpha_{i+2})  , \frac{1}{6} \right), \, 2<i<A-2\\
&\implies \pi(\bfalpha) \propto \exp\left(\frac{-1}{2} \sum_{i=3}^{A} (\alpha_{i} - 2\alpha_{i-1} + \alpha_{i-2})^2 \right) \\
&\implies \Delta^2 \alpha_i  \sim N(0, \mathcal{I}), \, i = 3, \dots, A.
\end{align*}
Hence the multivariate GMRF priors discussed earlier can also be viewed as matrix-normal priors on second differences (or first difference if the GMRF is a RW-1 model). There are also overlaps in the cross-strata correlations considered: the correlation in \cite{riebler:etal:12} is the same as our exchangeable model. 

The advantage of specifying the model only in terms of parameters that 
are identifiable (as we do here) is that we do not mistakenly report contrasts
that are only identified based on implicit constraints. To investigate
$\exp(\beta_{jr_1} - \beta_{jr_2})$ against $j$, we must 
specify $b_{r_1}$ and $b_{r_2}$, reminding us that the relative 
risk is only identifiable up to a multiplicative factor.
%

\section{Implementation} \label{S4}
\subsection{Model selection} \label{models}
We consider modifications from the full stratified APC model by allowing some parameters to be shared across regions.
We consider six models as summarized in Table \ref{modtab}. This is not an exhaustive list of all possible combinations of effects being shared or heterogeneous. Instead we prioritize sharing the baseline rates and age curvature 
with the rationale that age is a surrogate for biological processes that are less susceptible to environmental effects. We select a model using WAIC, which estimates the out of sample prediction error based on the log point-wise posterior predictive density of the observed data with a penalty to avoid over fitting \citep{gelman:etal:14}.

\begin{table}[h]
\centering
\caption{Summary of models to be considered.  A \checkmark means that parameter is constant across the strata.  For example, in M5, $\bfxi_{r0} = \bfxi_0$, but all of the other effects are allowed to vary.}\label{modtab}  
\begin{tabular}{|c|c|c|c|c|c|c|}
\hline 
 & M1 & M2 & M3 & M4 & M5 & M6 \\ 
\hline 
$\bfxi_0$ & \checkmark & \checkmark & \checkmark & \checkmark & \checkmark &  \\ 
\hline 
$\bfxi_\alpha$ & \checkmark & \checkmark & \checkmark & \checkmark &  &  \\
\hline 
$\bfxi_\beta$ & \checkmark & \checkmark &  & &  &  \\ 
\hline
$\bfxi_\gamma$ & \checkmark &  & \checkmark &  &  &  \\ 
\hline
\end{tabular}

\end{table}

For each of the models where the parameters are allowed to vary across strata, we also have a choice of correlation structure 
as outline in Section  \ref{sec:mvn}.  For simplicity we assume the same structure holds for each set of parameters, 
though we allow different hyper-parameter values.  For each model in Table \ref{modtab} (except M1), we consider three versions of the correlation structure: independent, exchangeable, and BYM2.  Thus the total number of models to choose from is 16.

\subsection{Software}
We have chosen to fit the hierarchical APC model with integrated nested 
Laplace approximations (INLA) as implemented in the \texttt{R-INLA} package 
\citep{rue:etal:09}.  INLA is a well-established method for fast and accurate approximations of the marginal 
posterior distributions for random effects and hyper-parameters when there are many latent Gaussian random effects, as is the case here.

Though we can view this model as a generalized linear mixed model with random slopes, there are some 
complications because we want to share the variance for certain blocks of the parameters.
We handle this using by passing in the relevant rows of $M$ via the $A$ matrix option in 
\texttt{control.predictor}, where the $M$ matrix itself is constructed via the \texttt{apc} package \citep{nielsen:16}. 
The Kronecker structured covariance is available using the 
\texttt{group} option. The full code for implementing all of the models considered in this paper is in the \hyperref[supps]{Supplementary Materials}.

\section{Application to the Human Mortality Database} \label{secHMD}

We apply this model to mortality rates in 25 European Union (EU) countries 
from the Human Mortality Database introduced in Section \ref{S1}. 
We fit our stratified APC models to these data from 1925-2015 and ages 0-80 
aggregating by 5-year age and period groups for a total of 17 age groups, 18 periods, and 34 
cohorts. For the BYM2 models, the concept of an adjacency is tricky because not all countries 
in Europe are covered in our example. We consider all countries that share a boundary to be 
adjacent but add in additional links along major rail, ferry, or road routes to create a single connected component. See the \hyperref[supps]{Supplementary Materials} for a map of the study region with addition links.  While it is not strictly necessary to create a single connected component, 
without these additional links we had four connected components, three of which only had two 
members (UK-Ireland, Sweden-Finland, Greece-Bulgaria). A priori the random effects for these  
disconnected pairs would not benefit from pooling of information across space, so we connected 
them to the remaining 19 areas.

\subsection{Hyper-prior distributions}
We follow the recommendations in \cite{smith:wakefield:16} by specifying  exponential priors on the precisions where the scale 
parameter is chosen based on a prior `confidence interval' for 
predictions in the age, period, or cohort effects. If we believe that 
the residual for predicting for the next period effect under the prior 
($\beta_{T+1} - \E[\beta_{T+1} \mid \beta_{1},\dots,\beta_{T}]$) to be 
no more than $\epsilon_\beta$ in absolute value $100\cdot(1-q)$\% of 
time, then this implies the rate in the hyper-prior for $\tau_\beta$ should 
be $(\epsilon_\beta/t_{1-q/2, \text{df} = 2})^2/2$. For the application 
in Section \ref{secHMD}, we take $q = 0.05$ and $\epsilon_\alpha = \log( 
1.2) = \epsilon_\beta = \log( 1.1)$, $\epsilon_\gamma = \log(1.01)$. For M6 (where the baseline rates are allowed to vary), we use $\epsilon_0= \log(1.05)$. 
In this example, we have more time periods than in the examples considered by \cite{smith:wakefield:16}, so we expect less sensitivity to these priors. 
For the correlation parameters, we use $\log((1+\rho(25 -1))/(1-\rho))\sim N(0,\sqrt{5})$ in the 
exchangeable case, and for the BYM2 model we use the penalized 
complexity prior from \cite{riebler:etal:16} assuming $Pr(\rho < 
0.5) = 0.5$. These are the default settings for these models in \texttt{R-INLA}.

For $\bfxi_0$, we use the baseline and two-slopes 
version from equation \ref{ACmid}. In this case, $U=9$ so the 
intercept is the log all-cause mortality rate for 40-45 year olds in 1925-1930 and the two slopes are the log relative risk between this group and the same cohort when aged 45-50 as well as the same initial group compared to the next cohort at the same age. In the case of all-cause mortality, we have substantial prior knowledge of the base-line rates. Thus we depart from the recommendation of 
\cite{smith:wakefield:16} and prefer to use an informative prior 
on $\bfnu_0$, the population averaged baseline effects:
$$\bfnu_0 \sim N( v = \{ \log(0.005), 0.3, -0.1\}, S = \text{diag}(1,0.1,0.1)).$$ 
This corresponds to a prior belief that the mortality of the baseline group is 5 per 1000 person years (95\% prior interval 0.7 to 35.5), a 35\% increase in risk for the age slope (27\% decrease to 151\% increase), and a 10\% decrease in risk for the cohort slope (51\% decrease to 68\% increase). 
These values are loosely based on data from Switzerland, which is in the HMD but not included in this example because Switzerland is not in the EU. For Switzerland the observed values are $\hat v = \{\log(0.0065), 0.256, -0.105\}$. In M6, $\bfnu_0$ is a matrix with identical columns.

One may question why the prior variances for $\bfnu_0$ are so small 
here.  We followed the suggestions of \cite{gabry:etal:2017} to
assess the multivariate properties of the prior specification by
checking that at least some simulations from the prior 
predictive distribution of $y$ were consistent 
with our expectations for real data.  With wider priors on $
\bfnu_0$, the simulated data were orders of magnitude larger than 
is sensible (e.g., everyone in the population would have to die multiple times). 
Using the prior above, we did not have this 
issue, and there is still plenty of 
mass in the prior on extreme log rates.
Under all 3 correlation 
structures, the maximum simulated count is greater than the 
observed maximum for the HMD data about 75\% of the time, and the 
simulated minimum is less than the observed minimum about 60\% of the time. 

\subsection{Results}
We begin by fitting all 16 models from Section \ref{models}.  The WAIC 
for each model is shown in Table \ref{waicres}.  The four non-spatial models allowing the age, period, and cohort curvatures to differ by location are clearly superior. The best fitting model according to WAIC allows all of the parameter blocks to vary by 
location (M6) with some pooling of information via an exchangeable correlation structure.

\begin{table}[h]
\centering
\caption{WAIC values for the 16 candidate models. A lower WAIC value indicates that a model has lower estimated out of sample prediction errors.} \label{waicres}
\begin{tabular}{rrrr}
  \hline
 & IID & Exchangeable & BYM2 \\ 
  \hline
M1 & 5107697.20 & -- & -- \\ 
  M2 & 5456311.16 & 5455501.60 & 6344619.42 \\ 
  M3 & 4223834.56 & 4223592.08 & 6595823.26 \\ 
  M4 & 3908640.41 & 3907772.54 & 6335372.21 \\ 
  M5 & 2882880.75 & 2881512.50 & 6265486.50 \\ 
  M6 & 2760168.21 & 2758382.19 & 6239374.61 \\ 
   \hline
\end{tabular}
\end{table}

\begin{figure}
\includegraphics[width = \textwidth, trim={1.7cm 0 1.7cm 0},clip]{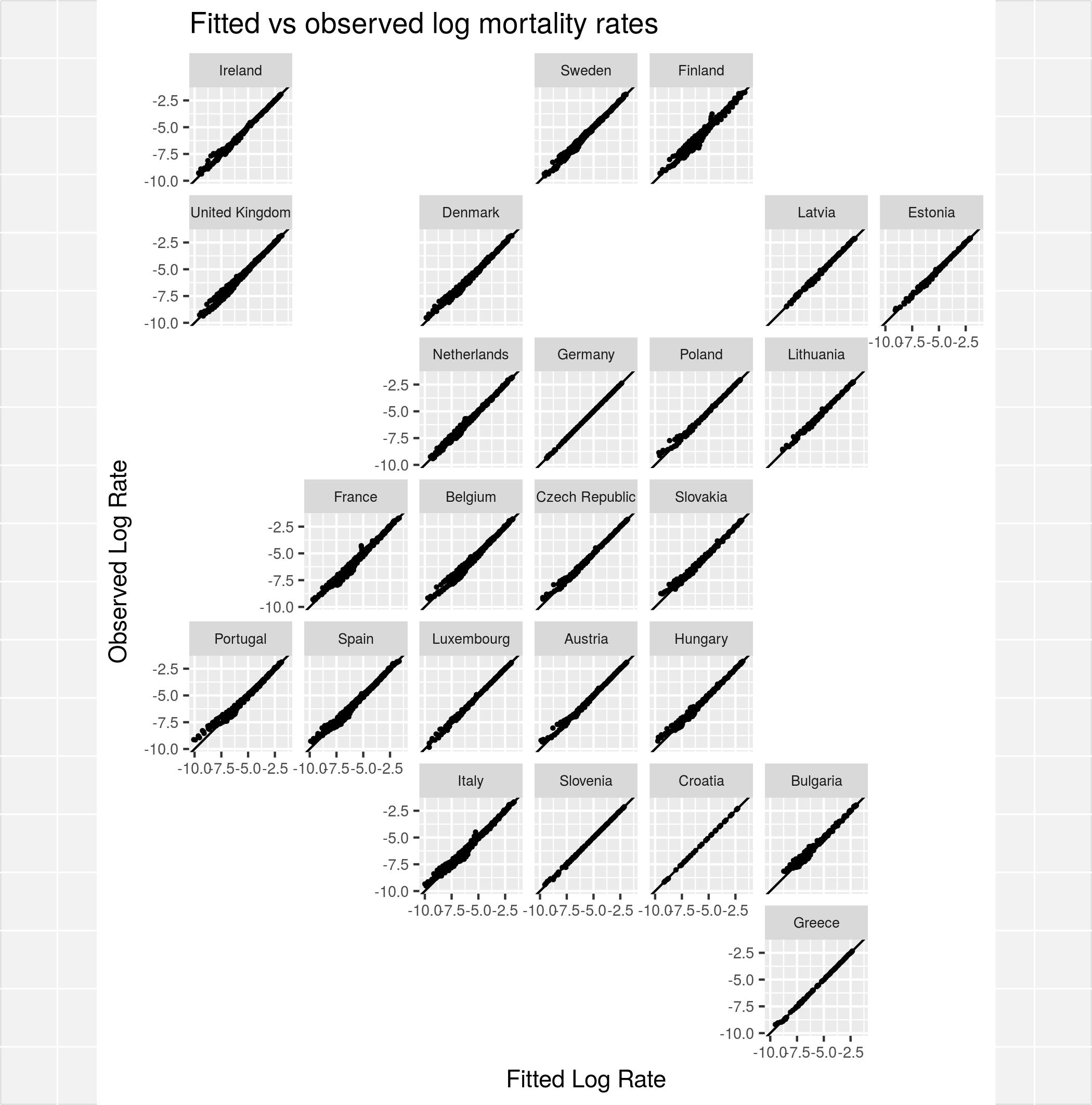}
\caption{Fitted vs observed log mortality rates in 25 EU member countries. The fitted rates are
based on the posterior medians from the M6-Exchangeable model.} \label{FvO}
\end{figure}

Table \ref{tab:ests} show the posterior medians along with upper and lower limits of 95\% credible intervals for the precisions (expressed as standard deviations), correlation parameters and mean baseline parameters ($\bfnu_0$) for the top four models. The estimates across the four models are broadly consistent.  The only major difference is the age slope $(\mu_{U+1,U} - \mu_{UU})$ is smaller when the baseline rate and slopes are allowed to vary. We also note that the width of the credible intervals is largest in M6-exchangeable model, which is expected given that this model has the largest number of hyper-parameters.  Figure \ref{FvO} shows the fitted (posterior medians) versus observed log rates for M6-exchangeable. There is some lack of fit in many countries for the smallest rates, and similar results were seen for the other three models.

\begin{table}[h]
\centering
\caption{Parameter (posterior medians and with upper and lower limits of 95\% credible intervals) for the top four models identified by WAIC.} \label{tab:ests}
\begin{tabular}{r|rr|rr|}
& \multicolumn{2}{c|}{Independent} & \multicolumn{2}{c|}{Exchangeable}\\
 &\multicolumn{1}{c}{M5} & \multicolumn{1}{c|}{M6}&\multicolumn{1}{c}{M5} & \multicolumn{1}{c|}{M6}\\ 
  \hline
 $\bfnu_0[1]$ & -4.95 (-4.96, -4.94) & -4.99 (-5.07, -4.91) & -4.95 (-4.96, -4.94) & -4.99 (-5.08, -4.91) \\ 
 $\bfnu_0[2]$ & 0.26 (0.26, 0.27) & 0.31 (0.25, 0.36) & 0.26 (0.26, 0.27) & 0.31 (0.24, 0.38) \\ 
  $\bfnu_0[3]$ & -0.089(-0.096,-0.083) & -0.091(-0.15,-0.034) & -0.088(-0.095,-0.081) & -0.088(-0.16,-0.018) \\ 
  \hline
    $\tau^{-1/2}_\alpha$ & 0.66 (0.62, 0.71) & 0.66 (0.62, 0.71) & 0.60 (0.45, 0.82) & 0.60 (0.45, 0.82) \\ 
   $\tau^{-1/2}_\beta$ & 0.095 (0.088, 0.1) & 0.091 (0.084, 0.098) & 0.10 (0.09, 0.12) & 0.10 (0.086, 0.12) \\ 
   $\tau^{-1/2}_\gamma$ & 0.064 (0.060, 0.068) & 0.060 (0.057, 0.064) & 0.064 (0.060, 0.069) & 0.061 (0.056, 0.066) \\ 
   $\tau^{-1/2}_0$ &  & 0.12 (0.10, 0.15) &  & 0.12 (0.096, 0.14) \\ 
  \hline
    $\rho_\alpha$ &  &  & 0.97 (0.94, 0.98) & 0.97 (0.95, 0.99) \\ 
    $\rho_\beta$ &  &  & 0.47 (0.32, 0.65) & 0.51 (0.36, 0.68) \\ 
    $\rho_\gamma$ &  &  & 0.20 (0.12, 0.31) & 0.22 (0.14, 0.34) \\ 
    $\rho_0$ &  &  &  & -0.024 (-0.041, 0.15) \\ 
   \hline
\end{tabular}
\end{table}

For the remainder of this section we concentrate on the estimates from the top model overall (M6-exchangeable).  In this model, the posterior median for the population-averaged mortality rate in the baseline group (40-45 year olds in 1925-1930) is $6.8$ death per 1000 person years (95\% CrI: $6.2$ -- $7.4$).  
The posterior median for the age slope, parameterized as the difference in log mortality between the baseline group and the same cohort when aged 45-50 is $0.31$, indicating a $36$\% increase in the morality rate  (95\% CrI: $27$\% -- $46$\%).  The posterior median for the cohort slope, parameterized as the difference in log mortality between the baseline group and the next cohort at the same age is $-0.088$, indicating an $8$\% decrease in risk  (95\% CrI: $15$\% -- $2$\%).  

\begin{figure}
\centering
\includegraphics[width = 0.99\textwidth]{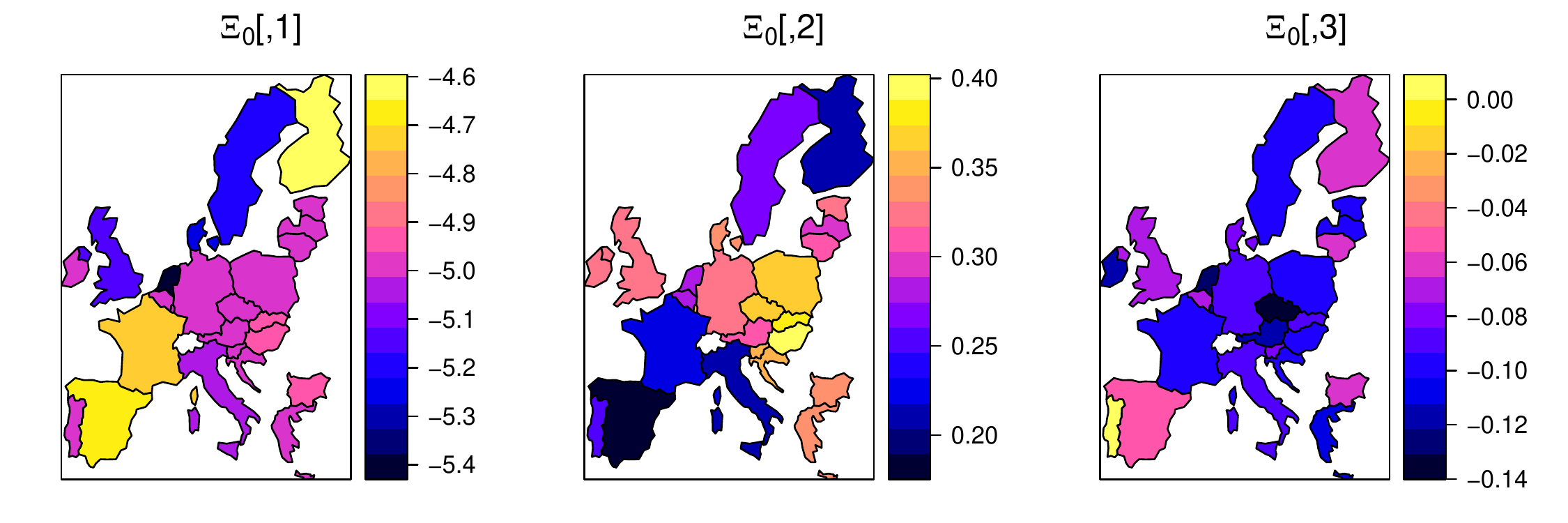}
\caption{Posterior medians of $\bfXi_0$ for the M6-Exchangeable model.} \label{fig:Xi0}
\end{figure}

The country-level baseline and slopes are moderately 
heterogeneous, but the deviations from the population averages are not significantly correlated.  
The range of posterior medians for the baseline rates is 
$4.6$ to $9.6$, with the highest risk in France, Spain 
and Finland and the lowest risk in the Netherlands (see 
Figure \ref{fig:Xi0}).  In eastern Europe, the 
deviations from the 
expected log baseline rate and slopes are the smallest, 
which is not surprising given that there is substantial 
missing data here in the early periods where $\{\mu_0^1, \mu_0^2, \mu_0^3\}$ are located.     

These three terms define a simple linear model for the time trend in each countries, and the remaining variance hyper-parameters describe the variance and correlation of non-linear deviations from this simple paradigm.  
The age curvature terms have the largest standard 
deviation ($0.60(0.45-0.82)$) and the largest correlation ($0.97(0.95-0.99)$).  This 
is not surprising because we know the relationship
between age and mortality usually has a non linear `J' shape. 
Further we expect the age effects to be
consistent across countries because they are surrogates
for processes that are less susceptible to
environmental differences.

\begin{figure}
\centering
\includegraphics[width = 0.99\textwidth]{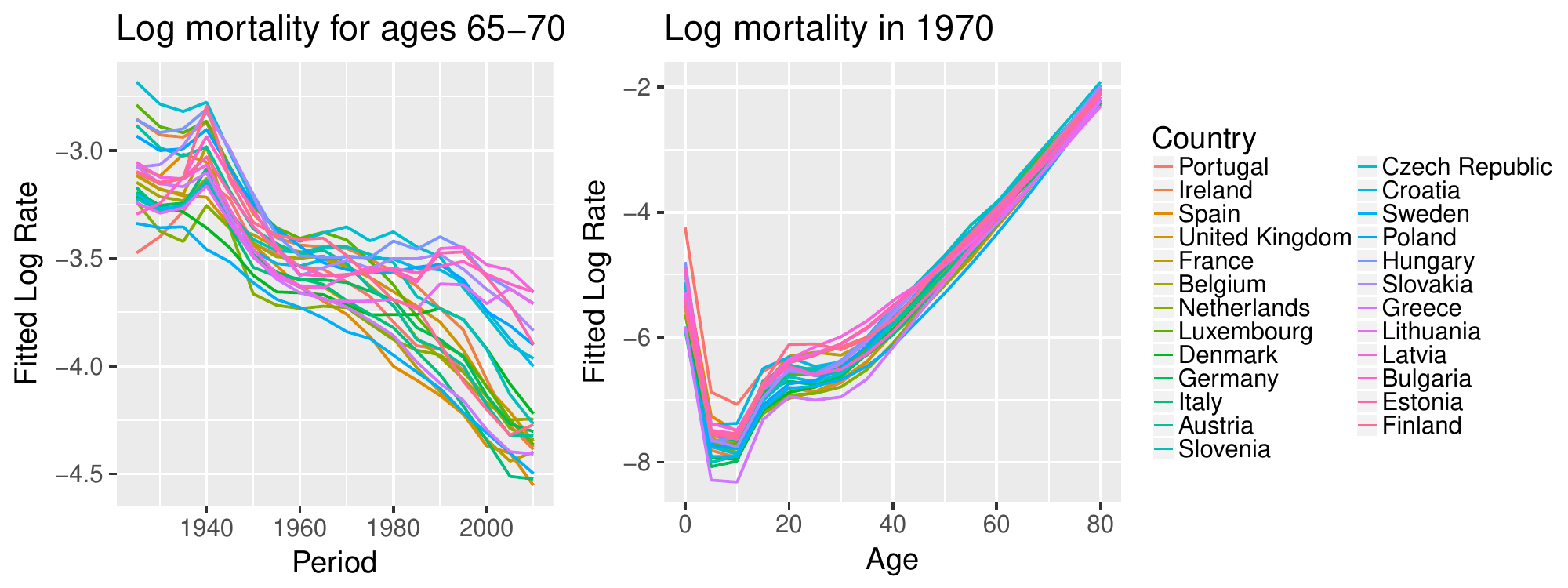}
\caption{Posterior medians of log rates in the 65-70 age group over period (left) and in 1970 across age groups (right) by country from the M6-exchangeable model.} \label{fig:CatDrift}
\end{figure}

We can get a sense of the age effects by plotting the 
fitted rates across the age groups for a given period 
(i.e., cross-sectionally) or for a given cohort (i.e., 
longitudinally).  We show the fitted cross-sectional age 
trends for 1970 (the middle time period) across all countries in Figure 
\ref{fig:CatDrift} (left).  The countries are ordered west 
to east by centroid.  We see the characteristic `J' 
shape and broadly the same curvature except that a few 
eastern European countries have somewhat straighter 
curves between 15 and 40.  It is important to remember
that differences in fit here are combinations of age and cohort effects and not solely due to differences in the age curvatures.

The period and cohort deviations are smaller and less 
correlated than the age curvatures.  Thus
we expect more heterogeneity in the trends of 
of disease along these two time scales than along
the age time scale.  
We can get a sense of the net effect of both the period 
and cohort effects, sometimes called drift, by 
plotting the fitted log rates for one age group across
time. Figure \ref{fig:CatDrift} (right) shows this plot for 
the 65-70 age group, across all countries. Pre 1960 the 
curves are very similar with a sharp peak in 1940 (WWII) 
and then steadily decreasing risk. The homogeneity in 
these earl years likely arises from the extent of 
missing data rather than genuine similarity. There is 
continued improvement in western Europe from 1960 
onward which does not seem to start in eastern Europe 
until 1985. Once again we remind the reader that 
differences in fitted rates seen here are the net effects
of period and cohort and not purely down to period effects.

\subsection{Hindcast for Germany 1960-1990}
As an additional check on the models, we imputed the 
mortality counts for Germany and compared against the 
true values reconstructed by combining the counts for 
East Germany (GDR) and West Germany (FRG). The HMD has data on 
unified Germany from 1990, which is the 
second shortest data record in this example after Croatia.
However, there are separate records for the GDR and FRG in the HMD, with the GDR records going back to 1960.

While the WIAC used in the previous section gives an 
estimate of the out of sample prediction, comparing
hindcasts of Germany mortality rates from 1960-1990
gives us an additional opportunity to test the predictive power of our models. For each of the M6 models, we sample from 
the posterior distribution of the log rates using the  
\texttt{inla.posterior.sample} function and then sample 
the number of deaths from the Poisson distribution.
We use the denominators ($N$'s) reconstructed from the HMD 
in this process, which may not be available in general.

\begin{figure}
\centering
\includegraphics[width = 0.99\textwidth]{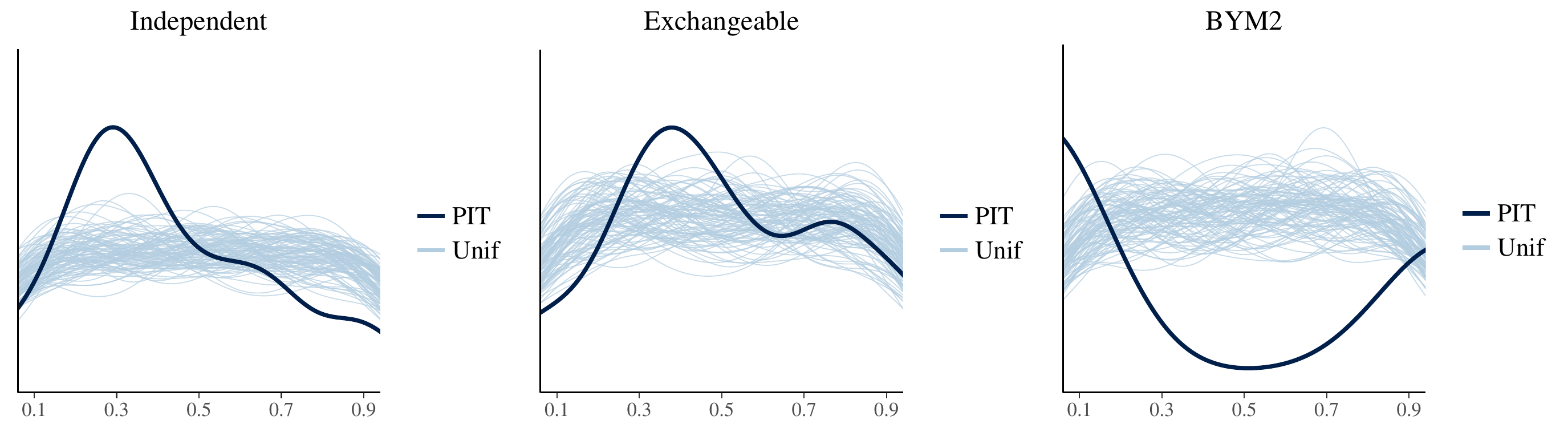}
\caption{Smoothed PIT histograms and histograms of simulated uniforms for M6 under each correlation model.} \label{fig:PIT}
\end{figure}

One way to assess these predictions is using probability integral transform (PIT)
histograms \citep{held:etal:2010}. For each age-period combination, we evaluate the empirical cumulative distribution function of the posterior samples at the observed count. If the posterior distribution for the missing data is a well calibrated prediction, then these proportions are approximately uniformly distributed. 
Figure \ref{fig:PIT} shows smoothed PIT histograms versus 
smoothed histograms for random samples from $\mathsf{U}
(0,1)$ for each correlation structure. 
There is miss calibration in all three models, but the exchangeable
model is clearly better calibrated than the other models. 
The shape of the BYM2 histogram also gives us some insight into why 
the spatial models performed so poorly in earlier comparisons. 
The $\cup$ shape means the posterior predictive distributions are
under-dispersed (too narrow).

\section{Discussion}
This article presents a Bayesian stratified APC model 
that allows for pooling of information on non-linear aspects of the 
age, period, and cohort effects across strata using matrix normal 
priors with separable covariance structures. We considered several 
cross-correlation models and put forward tools for selecting 
between more parsimonious versions of the full model.  
This approach is based on a fully identifiable parameterization, 
thus avoiding the major drawback of stratified APC models based on 
multivariate GMRFs. 

To our knowledge, the only other developments towards a 
multivariate APC model outside the GMRF framework are from
\cite{chernyavskiy:etal:2017}, who allow for
heterogeneity in the intercept and two trends but not the
curvature terms as we did here.  There are however numerous
age-space-time or age-strata-time models where only period effects are included.  For example in a recent paper \cite{goicoa:etal:2017} use splines to capture age-time trends in mortality and (in the most complicated model considered by the authors) there is a common age-time response surface along with 
region-specific age-time surfaces. Because there are only two time scales in an age-space-time model, we can directly interpret the age-time effects;  whereas, in the model proposed in this paper, we can only interpret the accelerations along these scales. 

In the application to the HMD data, the 
independent and exchangeable models with heterogeneous 
period curvatures only (M2) were superior to the models with  
heterogeneous cohort curvatures only (M3), suggesting 
that if one of the three time scales had to be dropped in favor of a more interpretable model, it should be cohort.  However, the model 
with heterogeneity in both period and curvature (M4) improved on 
both of these, indicating that a full APC model is appropriate.

Interestingly the spatial model in our application performed poorly 
compared to the exchangeable and independent cross-correlation 
models. One possible explanation for this is that the data are 
unbalanced in a spatially-structured way: the shortest data series 
are clustered in eastern Europe. Thus pooling information based on 
the neighborhood structure in the BYM2 model is not advantageous 
since there is little signal to draw on in the first place.
Understanding the role of spatially-structured missingness more generally is an avenue for future research.   

The parameterization on which we based our stratified model in Section \ref{S3}
is limited to data with equal-width age and time intervals. Relaxing this
assumption is an important direction for future development. Though 5-year age 
groups are standard, the time periods to which providers of official mortality
data aggregate differ considerably with data sometime provided at shorter (1-year) or longer (10-year) intervals. 
Even if all agencies report on 5-year time periods, the intervals may be misaligned.  
In spatial misalignment problems, one approach is to specify a model on smallest possible
units (e.g., single age-year intervals here) and then 
aggregate the mean as appropriate \citep{gelfand:2010}. 
If we follow this suggestion for the stratified APC model, the mean for the 
observed counts would no longer be log linear in the random effects. 
This poses additional computational challenges, and ascertaining 
which parameters remain identifiable is not straightforward.


\section*{Supplementary materials} \label{supps}
\begin{itemize}
\item Supplement A: Design matrix for APC models, directions for downloading data from the HMD and code for the analysis carried out Section \ref{secHMD}. (SupplementA.Rmd/.pdf)
\item Supplement B: folder of model scripts called in the .Rmd file (SupplementB.zip) 
\end{itemize}

\section*{Acknowledgments}
I thank the attendees of the Age-period-cohort 2 workshop (Nuffield College, Oxford, September 2017) for helpful discussions on the early stages of this work. 
{\it Conflict of Interest}: None declared.

\nocite{HMD}
\bibliographystyle{plainnat}
\bibliography{sapc.bib}

\end{document}